\documentclass[11pt,preprint]{aastex}
\pdfoutput=1
\usepackage{bm}
\usepackage{graphicx}
\usepackage{amsmath}    
\usepackage{amssymb}    
\usepackage{equation}
\usepackage{footnote}
\usepackage{color}
\usepackage{amsmath}
\usepackage{mathtools}
\usepackage{pdflscape}
\usepackage{multirow}

\let\oldfootsep=\footnotesep
\newcommand\ltsima{$\; \buildrel <\over\sim \;$}
\newcommand\simlt{\lower.5ex\hbox{\ltsima}}
\newcommand\gtsima{$\; \buildrel >\over\sim \;$}
\newcommand\simgt{\lower.5ex\hbox{\gtsima}}
\newcommand\etal{et~al.}

\newcommand\msun {M_\odot}

\newcommand\ie{{i.e.}}

\newcommand{\mathbold}[1]{\mbox{\boldmath $\bf#1$}}
\newcommand\piEbold{{\mathbold \pi_{\rm E}}}
\newcommand\mubold{{\mathbold \mu}}

\shorttitle{OGLE-2012-BLG-0563Lb is a Jupiter orbiting a K Dwarf}
\shortauthors{Bennett et al}

\begin{document}


\title{
Image-Constrained Modeling with
{\sl Hubble} and {\sl Keck} Images Reveals that OGLE-2012-BLG-0563Lb
is a Jupiter-Mass planet Orbiting a K Dwarf
}


\author{David~P.~Bennett\altaffilmark{1,2},
Aparna~Bhattacharya\altaffilmark{1,2},
Jean-Philippe~Beaulieu$^{3,4}$,
Naoki~Koshimoto$^{5}$,
Joshua~W.~Blackman\altaffilmark{6},
Ian~A.~Bond\altaffilmark{7},
Cl{\'e}ment~Ranc\altaffilmark{8},
Natalia~Rektsini\altaffilmark{3,4},
Sean K.~Terry\altaffilmark{1,2},
and
Aikaterini~Vandorou\altaffilmark{1,2},
 \\  } 
              

\affil{$^{1}$Code 667, NASA Goddard Space Flight Center, Greenbelt, MD 20771, USA;    \\ Email: {\tt bennettd@umd.edu}}
\affil{$^{2}$Department of Astronomy, University of Maryland, College Park, MD 20742, USA}
\affil{$^{3}$School of Physical Sciences, University of Tasmania, Private Bag 37 Hobart, Tasmania 7001 Australia}
\affil{$^{4}$Institut d'Astrophysique de Paris, 98 bis bd Arago, 75014 Paris, France}
\affil{$^{5}$Department of Earth and Space Science, Graduate School of Science, Osaka University, Toyonaka, Osaka 560-0043, Japan}
\affil{$^{6}$Physikalisches Institut, Universit\"{a}t Bern, Gessellschaftsstrasse 6, CH-3012 Bern, Switzerland}
\affil{$^{7}$Institute of Natural and Mathematical Sciences, Massey University, Auckland 0745, New Zealand}
\affil{$^{8}$Sorbonne Universitl{\'e}, CNRS, Institut d'Astrophysique de Paris, IAP, F-75014 Paris, France}

\begin{abstract}
We present high angular resolution imaging from the {\sl Hubble Space Telescope} 
combined with adaptive optics imaging results from the {\sl Keck}-II telescope to determine the
mass of the OGLE-2012-BLG-0563L host star and planet to be 
$M_{\rm host} = 0.801\pm 0.033\msun$ and $M_{\rm planet} = 1.116 \pm 0.087 M_{\rm Jupiter}$, respectively,
located at a distance of $D_L = 5.46\pm 0.56\,$kpc. There is a close-wide degeneracy in the light curve models
that indicates star-planet projected separation of $1.50\pm 0.16\,$AU for the close model and 
$8.41\pm 0.87\,$AU for the wide model.
We used the image-constrained modeling method to analyze the light curve data with constraints from this
high angular resolution image analysis. This revealed systematic errors in some of the ground-based light curve
photometry that led to an estimate of the angular Einstein Radius, $\theta_E$, that was too large by a factor of $\sim 2$.
The host star mass is a factor of 2.4 larger than the value presented in the \citet{fukui15} discovery paper. 
Although most systematic photometry errors seen in ground-based microlensing light curve photometry
will not be repeated in data from the {\sl Roman Space Telescope}'s Galactic Bulge Time Domain Survey,
we argue that image constrained modeling will be a valuable method to identify possible systematic
errors in {\sl Roman} photometry.
\end{abstract}


\section{Introduction}
\label{sec-intro}

The gravitational microlensing exoplanet detection method plays a key role in understanding the exoplanet
population in our Galaxy, despite the fact that the method was introduced only 33 years ago \citep{mao91}
and its sensitivity to low-mass planets was only recognized 28 years ago \citep{bennett96}. Microlensing's
unique sensitivity to low-mass planets in orbits wider than the Earth's orbit has led NASA to include a 
space-based microlensing exoplanet survey \citep{bennett02} as a key observing program that will use 
large fraction of the observing time for the {\sl Nancy Grace Roman Space Telescope} 
\citep{bennett_MPF,bennett18_wfirst,WFIRST_AFTA,penny19}.

A somewhat attractive feature of the microlensing method is that it is able to detect planets without detecting
any light from their host stars. This has enabled microlensing to detect the first Jupiter-like planet in
a Jupiter-like orbit around a white dwarf \citep{blackman-477}. It has also enabled the discovery of planets
with no evidence for any host star \citep{mroz_ffp18,mroz_2ffp,mroz_ffp20,ryu_ffp21,kim_ffp21,koshimoto23}, implying 
that they are likely to be unbound from any host star, although some could certainly be in wide orbits 
about host stars. Two of these candidate free-floating planets are likely to have masses slightly lower
than an Earth-mass \citep{mroz_ter_ffp,koshimoto23}, and a statistical analysis implies that these planets
are likely to be $6^{+6}_{-4}$ times more common than the known populations of bound planets
\citep{sumi23}.

However, the lack of clear detection of light from the planetary host star can be a problem for planetary
microlensing events with main sequence host stars. Methods of characterizing the physical properties of
planetary microlens systems, such as masses, distance, and orbital separation, 
have been discussed  by \citet{bennett07}. In some cases
it is possible to determine the masses and distances with light curve data alone if both the angular Einstein radius, 
$\theta_{\rm E}$, and the amplitude of microlensing parallax, $\pi_{\rm E}$, are both measured. However, this is
possible for only a fraction of events, usually with a bright source star and a duration long enough to see the
effects of the Earth's orbital motion in the light curve \citep[e.g.][]{muraki11} or observations by a telescope in a Heliocentric
orbit \citep[e.g.][]{street16}. 

Of the 29 events, with 30 planets, in the \citet{suzuki16} sample (hereafter S16), only three 
MOA-2009-BLG-266 \citep{muraki11}, MOA-2010-BLG-117 \citep{bennett-moa117}
and OGLE-2011-BLG-0265 \citep{skowron15}, have been characterized with mass measurements using only light curve
data. These three events provided  measurements of $\pi_{\rm E}$ and $\theta_{\rm E}$ that were precise enough to determine the
host star and planet masses. High angular resolution follow-up observations have been used to characterize almost all
of the other 26 S16 sample events. The only exceptions are 4 events with giant source stars, which make 
it very difficult to detect the host stars with high angular resolution follow-up images, although two of the giant source 
star events, MOA-2009-BLG-266 and OGLE-2011-BLG-0265 were characterized with $\pi_{\rm E}$ and $\theta_{\rm E}$ 
measurements. Excluding the events with giant source stars and the third event characterized with light curve
measurements  of $\pi_{\rm E}$ and $\theta_{\rm E}$ (MOA-2010-BLG-117), we are left with 22 events that we attempted
to characterize with high angular resolution observations with the the {\sl Hubble Space Telescope}
and/or {\sl Keck Telescope} adaptive optics (AO) imagers. Of these 22 events, we have detected the host stars for 13 events,
although in the case of the ``ambiguous" event of the S16 sample, the host star detection confirmed the non-planetary,
stellar binary model \citep{terry22}. The planetary host stars were detected for OGLE-2005-BLG-071,  
OGLE-2005-BLG-169, OGLE-2006-BLG-109, MOA-2007-BLG-192, MOA-2007-BLG-400, OGLE-2007-BLG-349,
MOA-2008-BLG-379, OGLE-2008-BLG-355, MOA-2009-BLG-319, MOA-2010-BLG-328, MOA-2011-BLG-262,
OGLE-2012-BLG-0950, and the event discussed in this paper, OGLE-2012-BLG-0563
\citep{batista15,bennett-ogle109,bennett15,bennett16,bennett-ogle71,bennett-moa379,aparna18,aparna21,gaudi-ogle109,terry21,terry-moa192}.
The papers on 3 of these events are still under preparation. There are also two events without host star detections 
despite very good high angular resolution follow-up data. \citep{aparna17} found that a candidate host star identified
in the discovery paper \citep{janczak10}, was actually a star that was unrelated to the microlensing event, and {\sl Keck}
images taken by \citet{blackman-477} ruled out main sequence stars, while microlensing parallax limits from
the light curve data \citep{bachelet12} ruled out brown dwarf, neutron star, and black hole hosts. This implies that
the host must be a white dwarf. Three of the remaining 7 S16 sample events have follow-up {\sl Hubble} imaging,
and all 7 have high quality follow-up {\sl Keck} AO imaging without any candidate host star detections. These will be used
to derive upper limits on main sequence planetary host stars.




The methods used to characterize exoplanet system are susceptible to systematic errors. Light curve features, 
like microlensing parallax, which are used to help determine lens system masses, are usually subtle and can be 
mimicked by the annual modulation of color-dependent atmospheric refraction \citep{terry-moa192}. They can
also be confused with astrophysical
effects, such as source star orbital motion (referred to as xallarap) or the orbital motion of the lens star and its planet.
Our methods to deal with these complications are detailed in \citet{bennett-moa379}, but the basic approach is to measure
and confirm the mass and distance determinations for the microlens planetary systems with multiple redundant methods.
The characterization of the
very first microlens exoplanet host star that was detected separating from its background source star, 
OGLE-2005-BLG-169L (\citep{bennett15,batista15}, was verified by observations from four different passbands,
using the $B$, $V$, and $I$ bands from {\sl Hubble} and the $K$ band from {\sl Keck}. Data from all four passbands yielded
consistent lens system mass measurements and were consistent with the lens-source separation predicted from the
light curve data. The relative lens-source proper motion between the 2011 {\sl Hubble} and 2013 {\sl Keck} images, also indicated
that the host star was at the location of the background source star at the time of the microlensing event.

This paper focuses on the analysis of planetary microlensing event OGLE-2012-BLG-0563 \citep{fukui15}, an event
in which high angular resolution follow-up observations in the $K$, $I$ and $V$ bands played an important role in
the discovery of a systematic error in some of the light curve photometry that had a significant effect on the inferred properties
of the OGLE-2012-BLG-0563L planetary system. In section~\ref{sec-event}, we discuss the characteristics of this microlensing
event and we review the findings presented in the discovery paper, which included AO observations with
the {\sl Subaru} telescope, when the microlensing event was in progress. This section also reviews the findings of the
companion paper by our group that describes the 2018 AO observations of this system with the {\sl Keck 2}
telescope \citep{aparna24}. Section~\ref{sec-HST} presents our analysis of our {\sl Hubble} observations of this target,
which suggests possible systematic errors in the light curve photometry, and our investigation of these systematic
errors is discussed in section~\ref{sec-syserror}. We present properties of the OGLE-2012-BLG--563Lb planetary microlens 
system in section~\ref{sec-lens_prop} and our conclusions in section~\ref{sec-conclude}.

\section{Planetary Discovery Paper and Adaptive Optics Observations of OGLE-2012-BLG-0563}
\label{sec-event}

The microlensing event, OGLE-2012-BLG-0563, was independently discovered by the OGLE and MOA collaborations,
as described in the \citet{fukui15} discovery paper (hereafter F15). Because it was predicted to reach high magnification, which implies 
a high sensitivity to planetary signals \citep{griest98,rhie00}, the $\,u$FUN and RoboNet groups began observations
more that 20 hours prior to peak magnification. The MOA group also added observations with the 0.61m
B\&C telescope at MJUO in the $I$-band filter at the Mt.\ John University Observatory (MJUO) in New Zealand on the
night of peak magnification. The RoboNet group observed the event from the Faulkes Telescope South (FTS) 2.0m 
telescope at Siding Springs Observatory, the Faulkes Telescope North (FTN) in Hawaii, and the Liverpool Telescope (LT)  in
the Canary Islands. However, the FTN and LT data do not cover enough of the light curve to constrain
the planetary models, so they are not included in any of our light curve modeling. The $\,u$FUN group followed this event
with the 1.3m SMARTS Telescope at CTIO in the $I$, $V$, and $H$ passbands, as a well as several amateur 
telescopes in New Zealand and Australia that were able to get good photometry at high magnification. We use $R$-band
data from the 0.4m Auckland Telescope and unfiltered data from the 0.3m PEST Telescope in Perth, Australia.

In addition to analyzing the light curve data, F15  also analyzed images of this target taken with the 
8.2m {\sl Subaru}  telescope on Mauna Kea in Hawaii using adaptive optics on 2012 July 28 (HJD = 2456137.8), when
the microlensing magnification had decreased to a factor of 1.3, according to our best fit model. These images had 
an angular resolution of $\sim 0.2^{\prime\prime}$, which is usually good enough to revolve the source from unrelated stars, 
but the lens-source separation at the time of these images was only $\sim 0.001^{\prime\prime}$, so the lens and source 
images could not be resolved into separate stellar images. 
Nevertheless, since the source brightness can usually be determined from the
light curve model, it is possible to detect excess stellar flux at the position of the source and lens stars. If there is excess flux
at the position of the source, there is a good chance that small or all of this excess flux is due to the lens star, which also 
hosts the planet. However, it is also possible that most of this excess flux could be due to a stellar companion to the 
lens or source star, or the chance superposition of a star that is unrelated to the microlensing event. These possibilities
can be tested with follow-up images of the microlens system that measure the separation and relative proper motion
of the source and candidate lens stars to determine if these are consistent with the microlensing 
model \citep[e.g.][]{bennett15,batista15} or not \citep{aparna17}. Since the lens-source separation for 
OGLE-2012-BLG-0563 was far too small to be measured at the time of the {\sl Subaru} 
observations, F15 consider the possibility that the excess flux is ``contaminated" by flux from a star other than
the lens star, such as a companion to the source or lens, or an unrelated star. They find that the implies host masses of 
$0.34^{+0.12}_{-0.20}\msun$, $0.19^{+0.21}_{-0.07}\msun$, and $0.13^{+0.17}_{-0.02}\msun$ for contamination fractions of 
0-10\%, 10-30\%, and 30-50\%, respectively.

\begin{figure}
\epsscale{0.9}
\plotone{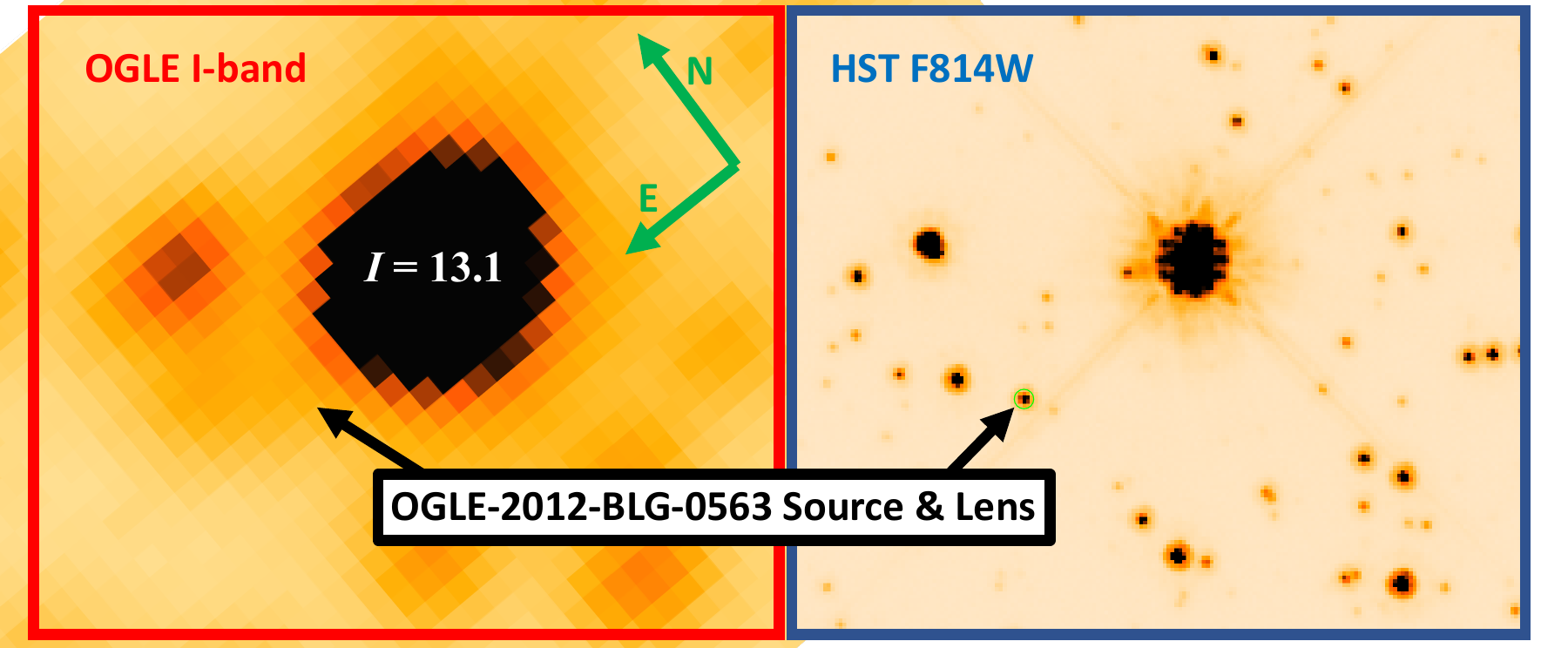}
\caption{(a) Comparison of the OGLE $I$-band reference frame image at the location of 
microlensing event OGLE-2012-BLG-0563 with the {\sl Hubble} image of the same area of the
sky. The source star with $I_s = 20.07$  is not visible in the OGLE image, but a saturated star
with $I = 13.10$ is very apparent at a separation of $2.2^{\prime\prime}$ is clearly visible \citep{fukui15}.
The blended source plus lens stars are clearly visible in the {\sl Hubble} image. The OGLE image was selected as the
reference frame because it has better seeing than most other OGLE images, and it has
much better seeing than most of the images used for light curve photometry. This bright star
contributes to systematic light curve photometry errors.
\label{fig-OGLE_v_HST}}
\end{figure}

F15 also note that the light curve photometry for this microlensing event is challenging due to the presence of 
a bright $I = 13.1$ star that is $2.2^{\prime\prime}$ from the microlensing event. This location of the microlensed 
source star is shown in OGLE $I$-band
and {\sl Hubble} $I$-band (WFC3/UVIS/F814W) images in Figure~\ref{fig-OGLE_v_HST}. F15 state that 
``the photometry was carefully done to minimize systematics" due to this bright neighbor star.
This star is saturated in most of the MOA images and in the best seeing OGLE images, and it is
likely that systematic errors persist in the OGLE-2012-BLG-0653 light curve photometry, despite the
best efforts of F15 to mitigate them.

One notable feature from the F15 analysis of this event is the relatively large angular Einstein radius,
which was determined to be $\theta_E = 1.36^{+0.14}_{-0.12}\,$mas. This is the second largest $\theta_E$
value out of the 27 events with $\theta_E$ measurements in the S16 sample, after the two-planet event
OGLE-2006-BLG-109 \citep{gaudi-ogle109,bennett-ogle109}. This large  $\theta_E$ value contributes to the
relatively small lens distance, $D_L = 1.3^{+0.6}_{-0.8}\,$kpc that they derive, with the minimal contamination fraction
of 0-10\%. (The inferred $D_L$ values are even smaller than this for larger assumed contamination fractions.)
However, the most prescient section of \citep{penny16} (section 6.2) pointed out that there was an excess
of planets with claimed distances of $D_L < 2\,$kpc among the microlens planetary events that had been published
at that time. There were six such events (OGLE-2006-BLG-109, MOA-2007-BLG-192, MOA-2010-BLG-328,
OGLE-2012-BLG-0563, OGLE-2013-BLG-0341, and OGLE-2013-BLG-0723) 
whereas their simulations predicted only one, under the assumption
that the planetary occurrence rate is independent of $D_L$. In fact, many of these planetary events with claimed
$D_L < 2\,$kpc have now been shown to be wrong. Event OGLE-2013-BLG-0723 was shown to have a very large
microlensing parallax signal that was due to the use of an incorrect model \citep{han-ob130723}. The event is due to
a stellar binary with no planetary signal. It was the attempt to fit the light curve to a binary star system with a planetary
signal that drove the microlensing parallax parameter to an unusually large value. With the correct model, the
inferred lens distance is $D_L = 3.1\pm 0.6\,$kpc. Event MOA-2007-BLG-192 was also found to have an unusually
large microlensing parallax value caused by a systematic error in the MOA data. Color dependent atmospheric
refraction can generate systematic errors in difference imaging photometry because stars of different colors will
have their positions shifted by different amounts, and this is especially significant in MOA data because of the 
unusually wide MOA-Red passband. This can induce seasonal systematic errors because color dependent 
centroid shifts are in different directions during the beginning and end of the Galactic bulge observing season.
Detrending methods to remove these systematic errors \citep{bennett12,bond17} have reduced both the 
microlensing parallax amplitude and the resulting lens system distance is $D_L = 2.16 \pm 0.30\,$kpc using
detections of the host star in follow-up observations with both {\sl Hubble} and {\sl Keck} AO images
\citep{terry-moa192}.

The modeling of MOA-2010-BLG-328 is perhaps the most complicated. The discovery paper \citep{moa328}
presented two alternative models: one with a strong microlensing parallax signal and planetary orbital motion and
one with a strong xallarap (source orbital motion) signal. In fact, the high angular resolution follow-up images
indicate that all of these light curve features are needed, as well as the magnification of the source companion
Vandorou \etal\ (in preparation). The final analysis for MOA-2010-BLG-328 reveals a source  distance 
$D_L = 3.19\pm 0.93\,$kpc, which is significantly larger than the distance of $0.81 \pm 0.10\,$kpc, reported in
the discovery paper \citep{moa328}.

In this paper, we show that the small $D_L$ value inferred by \citep{fukui15} for the OGLE-2012-BLG-0563 lens system
is due to a different type of systematic error, and the resulting lens system $D_L = 5.46\pm 0.55\,$kpc. This leaves the
two planet system OGLE-2006-BLG-109L and the planet in a binary system OGLE-2013-BLG-0341L \citep{gould14}
as the two remaining planetary events in the sample considered by \citet{penny16} at $D_L < 2\,$kpc. However, the
OGLE-2013-BLG-0341 event also had a large microlensing parallax signal and a linear trend that seemed to be
explained by the proper motion of a nearby star. However, OGLE data taken after publication failed to confirm 
this model to explain the linear trend, so it is unclear if the \citet{gould14} microlensing parallax signal is contaminated
by a systematic error. Thus, of the six potentially suspect events with $D_L < 2\,$kpc from \citet{penny16}, four have been
shown to be wrong, one is uncertain, and only OGLE-2006-BLG-109L remains at $D_L < 2\,$kpc. This agrees with the
conclusion of \citet{penny16} that ``there is very little chance that the [OGLE-2006-BLG-109] distance estimate is
significantly in error\rlap."

We have applied the lessons from these other analyses to our analysis of this event. In particular, we have 
used the same detrending methods \citep{bennett12,bond17} that were applied to the MOA-2007-BLG-192 follow-up imaging analysis
\citep{terry-moa192} to this event. However, we found that the MOA data preferred a much fainter source and
longer Einstein radius crossing time ($t_{\rm E}$) than the OGLE data. This is not very surprising because very
high magnification events are subject to a significant blending degeneracy \citep{alard97,distefano95}, because 
light curves with different $t_{\rm E}$ and source magnitudes can only be distinguished by the low magnification
parts of their light curves. The OGLE images have the benefits of a narrower passband, better seeing and a smaller angular 
pixel scale ($0.26^{\prime\prime}$ instead of $0.58^{\prime\prime}$). The MOA pixels subtend an angle that is 
a factor of 2.2 larger than the OGLE pixels, but the typical seeing in MOA images is only about a factor of 1.5 larger than
the typical OGLE seeing. This means that starlight in the OGLE images is spread over more pixels than in the MOA
images, and the wider MOA-red passband also increases the number of photons detected per pixel in the MOA images,
compared to OGLE. As a result, this bright neighbor star is saturated in a relatively small fraction of OGLE images, while it
is saturated in most MOA images. Therefore, we have decided to exclude the MOA images taken more than five days
before or after peak magnification at $ t = 6068.07\,$days in order to reduce the systematic photometry errors due to 
this bright neighbor star.

\subsection{Keck Adaptive Optics Follow-up Imaging}
\label{sec-Keck}

\begin{figure}
\epsscale{0.9}
\plotone{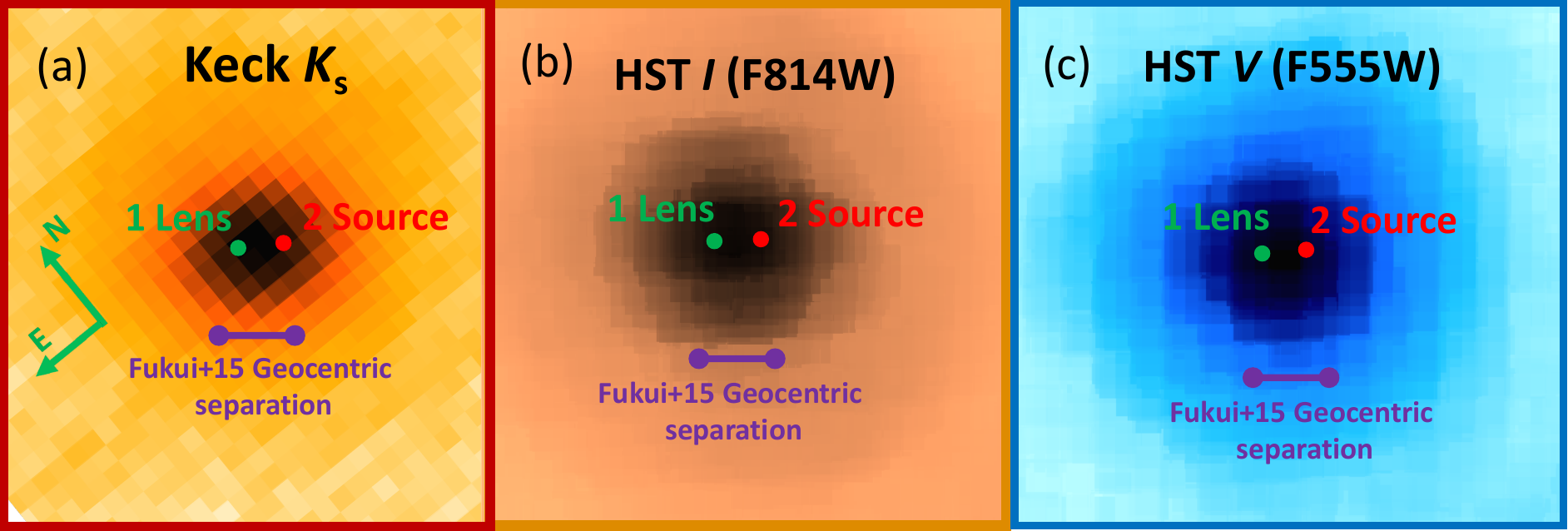}
\plotone{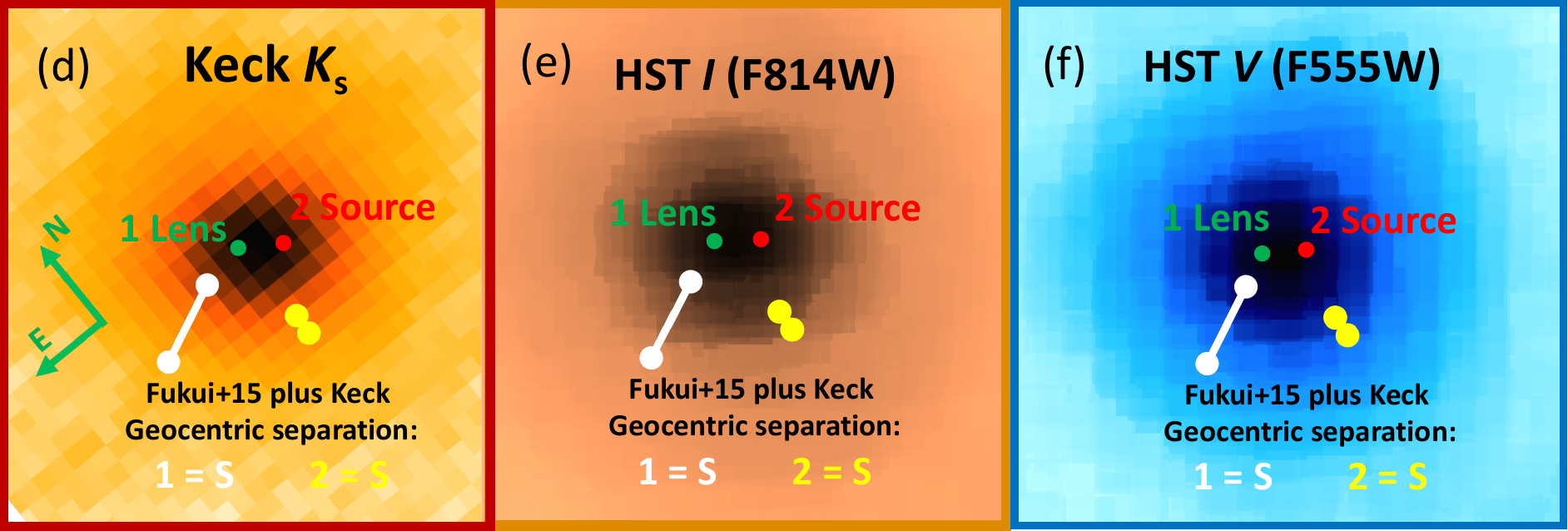}
\caption{Close up Keck $K$-band (panels a and d), and {\sl Hubble} $I$ (b and e) and $V$-band (c and f) images reveal
lens-source separations of $24.62\pm 0.67\,$mas of the blended lens+source stars in green and red. The $K$-band
data do not definitively indicate which is the lens and which is the source, and separation in the heliocentric frame 
measured by {\sl Keck} is smaller than the separation predicted by \citet{fukui15} in the geocentric frame. The conversion
of the {\sl Keck}  measurement to this geocentric frame depends on the lens distance and on which star is the source. 
The white and yellow lines
in the figure indicate {\sl Keck} predictions for the lens-source separation in the geocentric frame for assumptions that
the source is star 1 and 2, respectively. The predicted geocentric separations are $43.0\,$mas, and $10.7\,$mas
if stars 1 and 2, respectively, are the source stars. The \citet{fukui15} predicted geocentric separation is 
$38.4^{+3.6}_{-3.0}\,$mas, which is within $2\sigma$ of the {\sl Keck} result if star 1 is the source, but $9\sigma$ from
the {\sl Keck} result if star 2 is the source.
}
\label{fig-KIV_sep}
\end{figure}

The analysis of {\sl Keck} AO imaging of this target is presented in the \citet{aparna24} companion paper. The analysis
presented in this paper was based on 20 {\sl Keck} NIRC2 images with a combined
point-spread function (PSF) full-width half-max (FWHM) of
$63\,$mas. These images indicated two blended stars with a separation of $11.85 \pm1.38\,$mas and $20.05 \pm 0.80\,$mas 
in the North and East directions, respectively. The separation vector has a length of $23.29 \pm0.98\,$mas, which is
only 37\% of the FWHM, so the apparent lens and source stars are only partially resolved in these {\sl Keck} AO images,
as illustrated in panels (a) and (d) of Figure~\ref{fig-KIV_sep}.

Because of the strong overlap between the PSFs, the 2-star PSF models can trade flux between the two stars, resulting 
in a correlated uncertainty in their magnitudes. This analysis indicates that the combined brightness of the two stars is
$K_{LS} = 17.01\pm 0.05$, where we use $K$ to refer to the NIRC2 and 2MASS $K_s$ bands in this paper. The two-star
fit to the blended source and lens stars yields a magnitude difference of 
$K_1 - K_2= -0.075 \pm 0.178$, where $K_1$ and $K_S$
refer to the stars toward the North-East and South-West, respectively. In section~\ref{sec-HST}, we will show that
the North-East star is the lens (and planetary host) star and South-West star is the source star. The inferred magnitudes of these 
stars are $K_1 = 17.786 \pm 0.185$ and $K_2 = 17.861 \pm 0.185$, with a strong correlation between these magnitude 
uncertainties. F15 found $K_{LS} = 17.071 \pm 0.044$, which is within $1\sigma$ of our value (if we combine
the error bars), but their analysis is a bit more complicated because they had to correct for the microlensing magnification
at the time of their Subaru observations. This analysis led to a source magnitude of $K_{S\rm , Fukui} = 18.40\pm 0.10$,
which involved a transformation from a light curve measurement in the $H$ band. Our removal of the low-magnification
MOA data from microlensing light curve modeling of 
this event allowed the higher quality OGLE data to control the measured angular Einstein crossing
time, $t_{\rm E}$, which implies a brighter source star. However, F15 find that the excess flux blended with the 
source star gives $K_{\rm excess} = 17.69 \pm 0.11$, which is within $1\sigma$ of the $K$-band
magnitudes of both of the stars that were identified in the 2-star fit to the OGLE-2012-BLG-0563 {\sl Keck} images.

The results of our group's analysis of the {\sl Keck} images \citep{aparna24}
differ somewhat from the predictions of the OGLE-2012-BLG-0563Lb discovery paper
\citep{fukui15}. In particular, the  discover paper predicted a relative lens-source proper motion of 
$\mu_{\rm rel,G} = 6.4^{+0.6}_{-0.5}\,$mas/yr. Since the {\sl Keck} AO images were taken $6.00404\,$years
after the peak of the OGLE-2012-BLG-0563 microlensing event, the predicted lens-source separation was 
$\approx 6.4^{+0.6}_{-0.5} {\rm mas\,yr}^{-1} \times 6.00404\,{\rm yr} = 38.4^{+3.6}_{-3.0}\,$mas, but a two star
fit to the {\sl Keck} AO blended stellar image at the location of the OGLE-2012-BLG-0563 yielded a separation 
of $23.29 \pm0.98\,$mas. So the discovery paper's predicted separation from the light curve analysis is 65\% larger 
than the measured separation from the Keck observations.

The fact that separation of the stars in our 2-star fit to the blended source-plus-lens image is only 37\% of the
PSF FWHM means that it would be very difficult to detect a third star located between the other two.
Thus, it might be possible that the lens-source separation is consistent with the F15 prediction
of $38.4^{+3.6}_{-3.0}\,$mas if there was a third star located in between the lens and source stars.

In this case, however, there is an important difference between the coordinate systems used to determine the lens-source
relative proper motions, $\mu_{\rm rel}$. The light curve models can determine $\mu_{\rm rel}$ for events with finite source 
effects that determine the source radius crossing time, $t_*$. The angular source radius, $\theta_*$,
is determined from the extinction-corrected magnitude and color of the source 
\citep[e.g.][]{kervella_dwarf,boyajian14,adams18}, and then the amplitude of the lens-source relative proper motion is given by
\begin{equation} 
\mu_{\rm rel,G} = \theta_*/t_* \ ,
\label{eq-mu_relG}
\end{equation}
The subscript, G, in equation~\ref{eq-mu_relG} refers to the fact that this measurement or 
$\mu_{\rm rel}$  occurs in an inertial 
geocentric reference frame that moves with the Earth at the time of the peak of the microlensing event.
In contrast, the two-dimensional relative proper motion vector measured with high angular resolution follow-up
observations uses a heliocentric reference frame (plus a small contribution from geometric parallax that is almost
always insignificant). In most cases, the difference between the heliocentric lens-source relative proper motion
vector, $\mubold_{\rm rel,H}$, and the geocentric vector, $\mubold_{\rm rel,G}$ is relatively small, but this is not
the case for the ``best fit" parameters for OGLE-2012-BLG-563 reported by F15.

The relationship between the geocentric and heliocentric lens-source proper motions is given by
\citep{dong-ogle71}:
\begin{equation}
\mubold_{\rm rel,G} = \mubold_{\rm rel,H} - \frac{{\bm v}_{\oplus} \pi_{\rm rel}}{\rm AU}  \ ,
\label{eq-mu_helio}
\end{equation}
where ${\bm v}_{\oplus}$ is the projected velocity of the earth relative to the sun (perpendicular to the 
line-of-sight) at the time of peak magnification. This projected velocity is
${{\bm v}_{\oplus}}_{\rm E, N}$ = (25.1075, 1.4462) km/sec 
= (5.2963, 0.3051) AU/yr at the peak of the for OGLE-2012-BLG-563 microlensing light curve. This is close to the
Earth's full orbital speed of $28.8\,$km/sec because the event occurred near the middle of the Galactic bulge
season. So, the Earth's projected velocity term, ${\bm v}_{\oplus}$, in equation~\ref{eq-mu_helio} is larger 
than average, but the relative parallax, given by $\pi_{\rm rel} \equiv 1/D_L - 1/D_S$ would be very much larger 
than average if we adopted the small lens distance, $D_L = 1.3\,$kpc reported by F15. When we insert
the Earth's projected velocity and the definition of $\pi_{\rm rel}$ into equation~\ref{eq-mu_helio}, we have
\begin{equation} 
\mubold_{\rm rel,G} = \mubold_{\rm rel,H} - (5.2963,, 0.3051)\times (1/D_L - 1/D_S) \ ,
\label{eq-mu_helio2}
\end{equation}
Using the  F15 ``best fit" values for the source distance, $D_S = 9.1\,$kpc and $D_L = 1.3\,$kpc, we can find the convert the 
measured $\mubold_{\rm rel,H}$ to the geocentric coordinate system. However, the sign of the measured 
lens-source relative proper motion depends on which of the two stars is the source and which is the lens.
Assuming the F15 $D_L$ and $D_S$ values, which we argue are incorrect in Section~\ref{sec-syserror},
we find that the $\mubold_{\rm rel,G}$ vectors would be
\begin{equation}
\begin{eqalign}
         {\rm if\ Source = Star}\ 1\,:\ \ \mu_{\rm rel,G,E} =& -6.83\pm 0.13\,{\rm mas/yr},\ \ \mu_{\rm rel,G,N} = -2.17\pm 0.23\,{\rm mas/yr}\ ,\\
         {\rm if\ Source = Star}\ 2\,:\ \ \mu_{\rm rel,G,E} =& -0.15\pm 0.13\,{\rm mas/yr},\ \ \mu_{\rm rel,G,N} = 1.77\pm 0.23\,{\rm mas/yr} \ ,
\label{eq-keck2hst}
\end{eqalign}
\end{equation}
for these two choices of the source stars.
The light curve models only provide the length of the $\mubold_{\rm rel,G}$ vector, so these results would imply that 
the $\mubold_{\rm rel,G}$ have lengths of $\mu_{\rm rel,G} = 7.17\pm 0.15\,$mas/yr if star 1 is the source and 
$\mu_{\rm rel,G} = 1.78\pm 0.23\,$mas/yr if star 2 is the source. These vectors are displayed as white and yellow lines
in Figure~\ref{fig-KIV_sep} panels d, e and f. These give both the separation and the orientation of the two stars in the
geocentric frame, but the {\sl Keck} and {\sl Hubble} images are in the heliocentric frame. Since the light curve only provides
information on the magnitude of the  $\mubold_{\rm rel,G}$ vector, it is only the lengths of these vectors that should be compared
with the F15 geocentric separation prediction shown in Figure~\ref{fig-KIV_sep} panels a, b and c.
The F15 value of $\mu_{\rm rel,G} = 6.4^{+0.6}_{-0.5}\,$mas/yr, is only 1.3$\sigma$ smaller than the $\mu_{\rm rel,G}$
value for star 1 as the source, so this source identification would seem to be a better fit to the Keck data,
as long as we assume that the  F15 lens distance value of $D_S \simeq 1.3\,$kpc is correct. 
However, the direction of the $\mubold_{\rm rel,G}$ vector is identical to the direction of the $\piEbold$ vector and the 
length of the $\mubold_{\rm rel,H}$ vector is proportional to the angular Einstein radius, $\theta_{\rm E}$, which is inversely 
proportional to the source radius crossing time, $t_*$. Thus, constraints on $\mubold_{\rm rel,H}$ also modify light 
curve parameters. As we discuss in our companion paper
\citep{aparna24}, our image-constrained modeling \citep{bennett-moa379} of this event
raises some doubt that star 1 could be the source. When we apply the 
{\sl Keck} constraints on the $K$-band magnitude and $\mubold_{\rm rel,H}$ measurement under the assumption that
star 1 is the source, the fit $\chi^2$ rises 
$\chi^2$ by $\simeq 35$ compared to the best light curve model without these constraints.

We should also note that the uncertainties in the host star mass and distance estimates reported by  F15 are
relatively large. Also, the mass-luminosity relation for a close binary host star system would imply that the lens system
is more distant \citep{bennett16}, so the difference between $\mubold_{\rm rel,H}$ and $\mubold_{\rm rel,G}$ could be
much smaller than implied by the ``best fit" parameters reported by F15. So, the possibility that the 
lens-source separation measurement might be contaminated by a third star, as mentioned earlier in this section cannot be 
excluded on the basis of the Keck data.


\section{Hubble Space Telescope Follow-up Observations and Analysis}
\label{sec-HST}

\begin{deluxetable}{cccccccc}
\tablecaption{{\sl Keck} and {\sl Hubble} $\mubold_{\rm rel,H}$ and magnitude values \label{tab-mu-rel-mag}}
\tablewidth{0pt}
\tablehead{Passband & $\mu_{\rm rel,H,N}$ & $\mu_{\rm rel,H,E}$ & $\mu_{\rm rel,H}$ & Star 1 mag & Star 2 mag  \\
 & mas/yr & mas/yr & mas/yr & (lens) & (source) } 
\startdata
Keck $K$               & $1.974 \pm 0.229$ & $3.339 \pm 0.134$ & $3.879 \pm 0.164$ & $17.786\pm 0.185$ & $17.861\pm 0.185$ \\
HST F814W ($I$)  & $2.564 \pm 0.206$ & $3.508 \pm 0.194$ & $4.345 \pm 0.198$ & $19.988 \pm 0.106$ & $19.891\pm 0.096$  \\
HST F555W ($V$) & $2.261 \pm 0.277$ & $3.453 \pm 0.249$ & $4.128 \pm 0.258$ & $21.714\pm 0.123$ & $21.407\pm 0.100$   \\
\tableline
weighted mean      & $2.291 \pm 0.134$ & $3.403 \pm 0.101$ & $4.103 \pm 0.112$ &  &   \\
\tableline
\enddata\\
\vspace{0.15cm}
\parbox{16.1cm}{
The Heliocentric relative proper motion and magnitude measurements for stars 1 and 2. 
The relative proper motion refers to the motion of star 1 minus star 2. }
\end{deluxetable}

On 2018 May 26, we obtained a single orbit of {\sl Hubble} observations
from program GO-15455, using the WFC3/UVIS camera. This was 6.009191 years after the peak of the event, and just 
about 45 hours after the {\sl Keck} observations that were taken     6.004040 years after the peak of the even.
We obtained $16\times 85\,$sec.\ dithered exposures with the F814W
filter and $14\times 92\,$sec.\ dithered exposures with the F555W filter using the UVIS2-C1K1C-SUB aperture to minimize CTE
One image in each of these passbands was taken in the .UVIS2-2K2C-SUB aperture, but these were not used in the analysis.
(The {\sl Hubble} data used in this paper can be found in MAST: \dataset[10.17909/wd7f-vv60]{http://dx.doi.org/10.17909/wd7f-vv60}.)
The analysis was done with a modified version of the codes based on an early version of hst1pass \citep{hst1pass} used by
\citet{bennett15} and \citet{aparna18}. It differs from hst1pass  in that is has specialized routines to accurately measure the
photometry and astrometry of partially resolved stars, due coordinate transformations to {\sl Keck} images and to calibrate
photometry to the OGLE-III database \citep{ogle3-phot}. Like hst1pass, our {\sl Hubble} analysis code analyze the data 
from the original images without any resampling
in order to avoid the loss of resolution that the combination of dithered, undersampled images would provide.

The coordinate transformation between the {\sl Keck} and  {\sl Hubble}  images was done with 16 stars brighter than
$K < 14.0$, yielding the transformation 
\begin{equation}
\begin{eqalign}
         x_{\rm hst} =& -0.2001567\, x_{\rm keck}+ 0.1507597\, y_{\rm Keck}+549.1780\\
         y_{\rm hst} =& -0.1503665\, x_{\rm keck} -0.2010147\, y_{\rm Keck}+793.6765 \ ,
\label{eq-keck2hst}
\end{eqalign}
\end{equation}
from {\sl Keck} to {\sl Hubble} WFC3/UVIS pixels. The RMS scatter for this relation is 
$\sigma_x = 0.0224\,{\rm pix} = 0.896\,$mas and $\sigma_y = 0.0245\,{\rm pix} = 0.981\,$mas, since the
WFC3/UVIS pixels subtend 40\,mas. This separation is small because the time interval between the images
is $< 2\,$days, so the offset due to stellar proper motion is negligible.

\begin{figure}
\epsscale{0.66}
\plotone{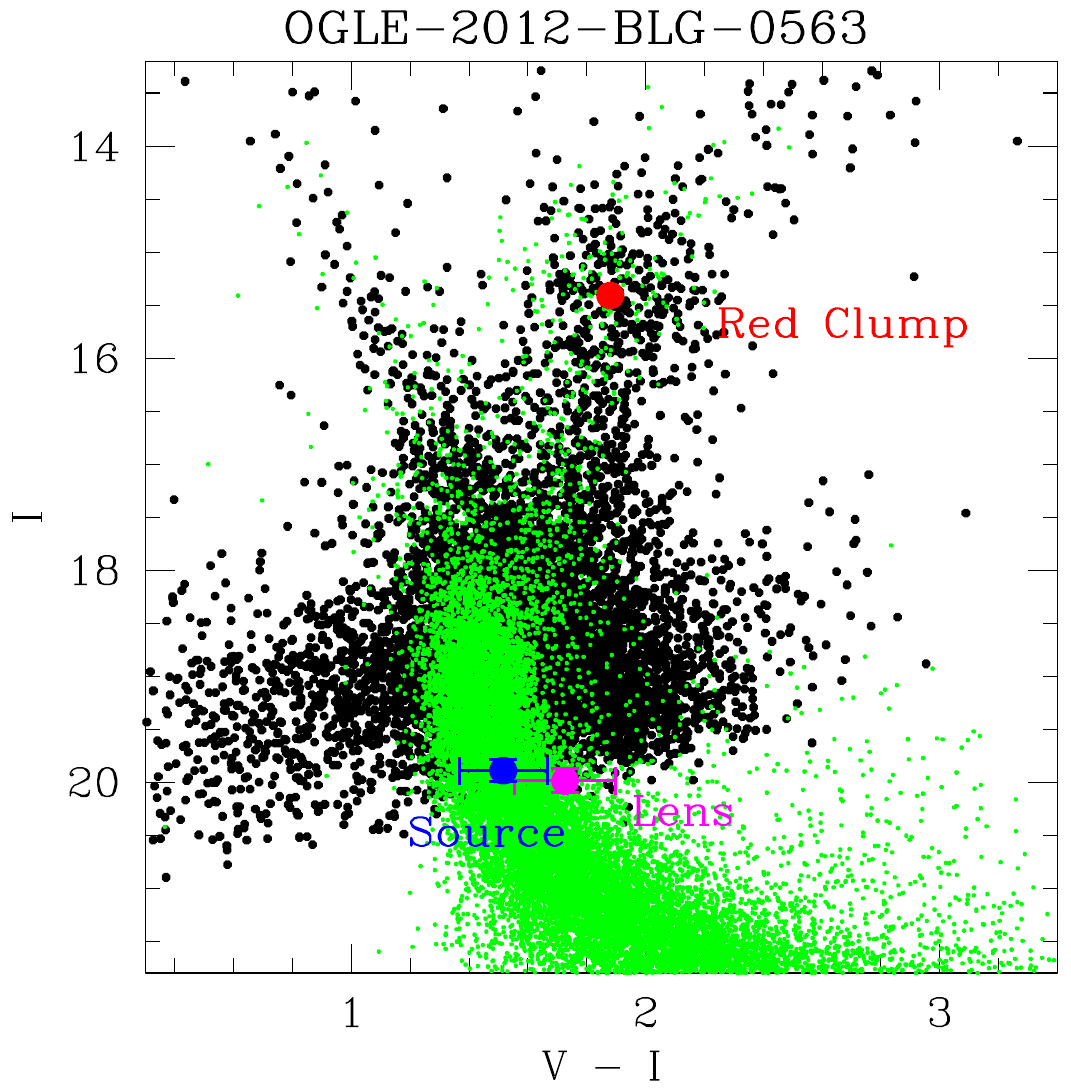}
\vspace{-0.35cm}
\caption{The CMD of OGLE-3 stars within 120 arc seconds of microlensing event OGLE-2012-BLG-0563  (black dots),
with the {\sl Hubble} CMD of Baade's Window (green dots) \citep{holtzman98}, transformed to the same extinction and Galactic bar distance as the OGLE-2012-BLG-0563 field. The
red spot is the red clump giant centroid; the source and lens magnitudes from our {\sl Hubble} observations
are indicated in blue and magenta. Since the lens star is likely
to be in the disk or the near side of the bulge, it is typically brighter than the bulge main sequence.
\label{fig-cmd}}
\end{figure}

The {\sl Hubble} photometry was calibrated to the OGLE-III catalog \citep{ogle3-phot} using 7
relatively bright OGLE-III stars stars in the $I$ band and 5 bright stars in the $V$ band. These stars were selected 
because they had no nearby neighbors that were close enough to contaminate the OGLE photometry. The RMS scatter of
these stars to the best fit photometry zero-point was 0.0265 magnitude in the $I$-band and 0.0285 magnitudes in the
$V$-band. We assume a calibration uncertainty of 0.04 magnitudes that is added in quadrature to the uncertainties in 
the instrumental magnitudes to get the results reported in Table~\ref{tab-mu-rel-mag}.

The locations of the stars 1 and 2 are shown as magenta and blue dots on the OGLE-III color magnitude diagram 
in Figure~\ref{fig-cmd}, where they are labeled as the lens and source stars, respectively based on the 
arguments given below. The green points in Figure~\ref{fig-cmd} are from the \citet{holtzman98}
Baade's Window color magnitude diagram, converted to the same extinction and Galactic bar distance
as the OGLE-2012-BLG-0563S source star. Star 1 is located on the red edge of the bulge main sequence,
as we would expect for an intrinsically fainter and slightly redder lens star in the foreground of the bulge

Following \cite{bennett14}, we determine the $V$ and $I$ band extinction from the stars 
$120^{\prime\prime}$ of the OGLE-2012-BLG-0563S source by measuring the centroid of the
red clump giant stars. We find $I_{\rm rcg} = 15.41 \pm 0.05$ and $(V-I)_{\rm rcg} = 1.88 \pm 0.030$,
Following \cite{nataf13}, we use an intrinsic red clump star magnitude and color of $I_{\rm rcg0} = 14.343$ 
and $(V-I)_{\rm rcg0} = 1.06$. This yields extinction
values of $A_I = 1.067\pm 0.050$ and $A_V = 1.887\pm 0.03$, implying a color excess of $E(V-I) = 0.82$. 

The coordinate transformation in equation~\ref{eq-keck2hst} enables us to put the relative proper motion measurements
($\mubold_{\rm rel,H}$) from {\sl Keck} and {\sl Hubble} on the same Heliocentric coordinate system, as shown 
in Table~\ref{tab-mu-rel-mag}, which also indicates the calibrated magnitudes for stars 1 and 2 in all three passbands
($K$, $I$, and $V$).  The $V-K$ colors of the two stars are $V_1-K_1 = 3.93 \pm 0.22$ and $V_2-K_2 = 3.55\pm 0.21$.
We use the OGLE-III photometry catalog \citep{ogle3-phot} and the \cite{nataf13} red clump
star color magnitude centroid values at the Galactic longitude of this event to find extinction values of
$A_I = 1.067\pm 0.03$ and $A_V = 1.887\pm 0.06$. These are similar to the values of 
$A_I = 1.02\pm 0.04$ and $A_V = 1.84\pm 0.06$ reported by
F15. We find a $K$-band extinction of $A_K = 0.157\pm 0.02$, from the \cite{surot20} value of the color 
excess at the OGLE-2012-BLG-0563 location using the \cite{nish06} infrared extinction law. So, the 
extinction corrected magnitude and colors for star 2 are $I_{2,0} = 17.82\pm 0.10$, $V_{2,0} - I_{2,0} =0.70\pm 0.15$ and 
$V_{2,0} - K_{2,0} = 1.82\pm 0.22$, which is consistent with a G-dwarf in the bulge, as inferred by
F15. For star 1, we find $I_{1,0} = 17.92\pm 0.10$, $V_{1,0} - I_{1,0} =0.91\pm 0.18$ and 
$V_{1,0} - K_{1,0} = 2.20\pm 0.23$, assuming the same extinction as the red clump centroid. 

We performed a preliminary light curve analysis using all the light curve data from F15, except the MOA data
was reduced by an improved MOA photometry code \citep{bond01,bond17} for both the custom MOA-red passband and the MOA-V 
passband. These were also calibrated to the  OGLE-III catalog \citep{ogle3-phot} . These results indicated preliminary
calibrated source star magnitudes of $I_S = 19.941\pm 0.042$ and $V_S = 21.373\pm 0.043$, which are consistent with the
identification of star 1 as the source.


These measurements do not match the conclusions of F15, who
reported a host star mass of $M_{\rm host} = 0.34^{+0.12}_{-0.20}\msun$, at a distance of 
$D_L = 1.3^{+0.6}_{-0.8}\,$kpc. A host star of $0.34\msun$ would have a color of $V-K = 6.45$ with
the same extinction as the source, or  $V-K = 5.58$ with only half the extinction of the source, which is still 1.65
magnitudes redder than star 1.  Even a star at the 2$\sigma$ mass upper limit or $M_{\rm host} = 0.58\msun$, 
would have $V-K=4.72$, which is $0.78\pm 0.22$ redder than star 1, assuming  half the extinction of the source star.
Thus, it seems unlikely that our high angular resolution follow-up observations will confirm the  F15
analysis.

\begin{deluxetable}{ccccc}
\tablecaption{Best Fit Model Comparison
                         \label{tab-fitcomp} }
\tablewidth{0pt}
\tablehead{
& \multicolumn{3}{c} {with FTS, B\&C data} & no FTS, B\&C data \\
\colhead{parameter}  & \colhead{unconstrained} & $K_L$ only & \colhead{constrainted} & \colhead{constrainted} 
}  
\startdata
$t_{\rm E}$ (days) &62.058  & 62.157 & 63.979 &  63.768 \\   
$t_0$ (${\rm HJD}^\prime$) &  6069.0290 & 6069.0291 & 6069.0275 & 6069.0281  \\
$u_0$ & -0.0017541  & - 0.0017550 & -0.0017192 & -0.0017281  \\
$s$ & 0.40159 & 0.40559 &0.41513 & 0.42452  \\
$\alpha$ (rad) & 0.49856  & 0.49769 & 0.49397 & 0.49231  \\
$q \times 10^{3}$  & 1.4891  & 1.4604 & 1.3866 & 1.3088    \\
$t_\ast$ (days) & {\bf  0.01951 } & {\bf 0.02488 } &  {\bf 0.04004 } & {\bf 0.04319}  \\
$\pi_{\rm E,N}$ &  0.59927 & 0.60754 &  0.05361 & 0.05821  \\
$\pi_{\rm E,E}$ &  0.11011  & 0.11019 & 0.08207  & 0.07955     \\
$D_{s}$ (kpc) & -  &8.0306  & 8.8438 & 8.4531  \\
FTS $\chi^2$ & 185.52 & 187.55 & 213.516 & - \\
B\&C $\chi^2$ & 528.29  & 529.84 & 542.155 & - \\
OGLE-$I$ $\chi^2$ & 670.04 & 667.80 & 662.199 & 660.094  \\
MOA-Red $\chi^2$ & 164.804 & 164.231 & 165.390 & 159.305   \\
light curve $\chi^2$ & 2064.65 & 2066.13 & 2100.75  &  1344.10  \\
dof &  $\sim 2169$ & $\sim 2169$ & $\sim 2178$ & $\sim 1460$   \\
$\chi^2$(no FTS,B\&C) & 1350.84 & 1348.74 & 1345.07 & 1344.10  \\
dof(no FTS,B\&C) & $\sim 1451$ & $\sim 1451$ & $\sim 1460$  & $\sim 1460$  \\
$D_L$ (kpc) &  -  & 1.154 & 5.586  & 5.424 \\
$M_{\rm host}/\msun$ & -  & 0.239 & 0.843  & 0.835 \\
$\theta_{\rm E}$ (mas) & {\bf 1.532}  & {\bf 1.201}  &{\bf  0.673} & {\bf 0.670} \\
\enddata \\
\vspace{0.15cm}
The $\pi_{\rm E}$ values use an inertial geocentric coordinate system that moves with 
the Earth at $t_{\rm fix} = 6069\,{\rm HJD}^\prime$. 
The image constraints increase source radius crossing time, $t_*$ and and 
decrease the implied angular Einstein radius, $\theta_{\rm E}$ (both highlighted in bold-face) by a factor of $\sim 2$.
\end{deluxetable}

In order to investigate this, we modeled this event with the image-constrained modeling version 
of the \texttt{eesunhong}\footnote{\url{https://github.com/golmschenk/eesunhong}} light curve
modeling code, described in detail in \citet{bennett-himag,bennett-moa379}. This code includes Gaussian constraints 
for seven of the measurements listed in Table~\ref{tab-mu-rel-mag}. These include the weighted mean values
of the North and East components of $\mubold_{\rm rel,H}$, using both the {\sl Keck} and {\sl Hubble} data,
the $K$, $I$, and $V$ host star magnitudes, and the $I$ and $V$ source star magnitudes. (Because we have no
light curve measurements in the $K$ band, we impose no constraints on the source $K$ magnitude.)
Also, because the separation of the lens and source stars is $< 40$\% of the FWHM of the {\sl Keck} images and
only 62\% of a {\sl Hubble} WFC3/UVIS pixel, the images of the lens and source stars have significant overlap,
and this results in a significant correlation between their magnitude measurements. A fainter magnitude for one 
star can be compensated by a brighter magnitude for the other. As a result, the magnitude of the combined 
lens plus source stars can be determined more precisely than the individual magnitudes. Therefore, we
impose two additional constraints on the combined lens and source magnitudes, $I_{LS} = 19.186 \pm 0.080$
and $V_{LS} = 20.797 \pm 0.080$.

\begin{figure}
\epsscale{0.9}
\plotone{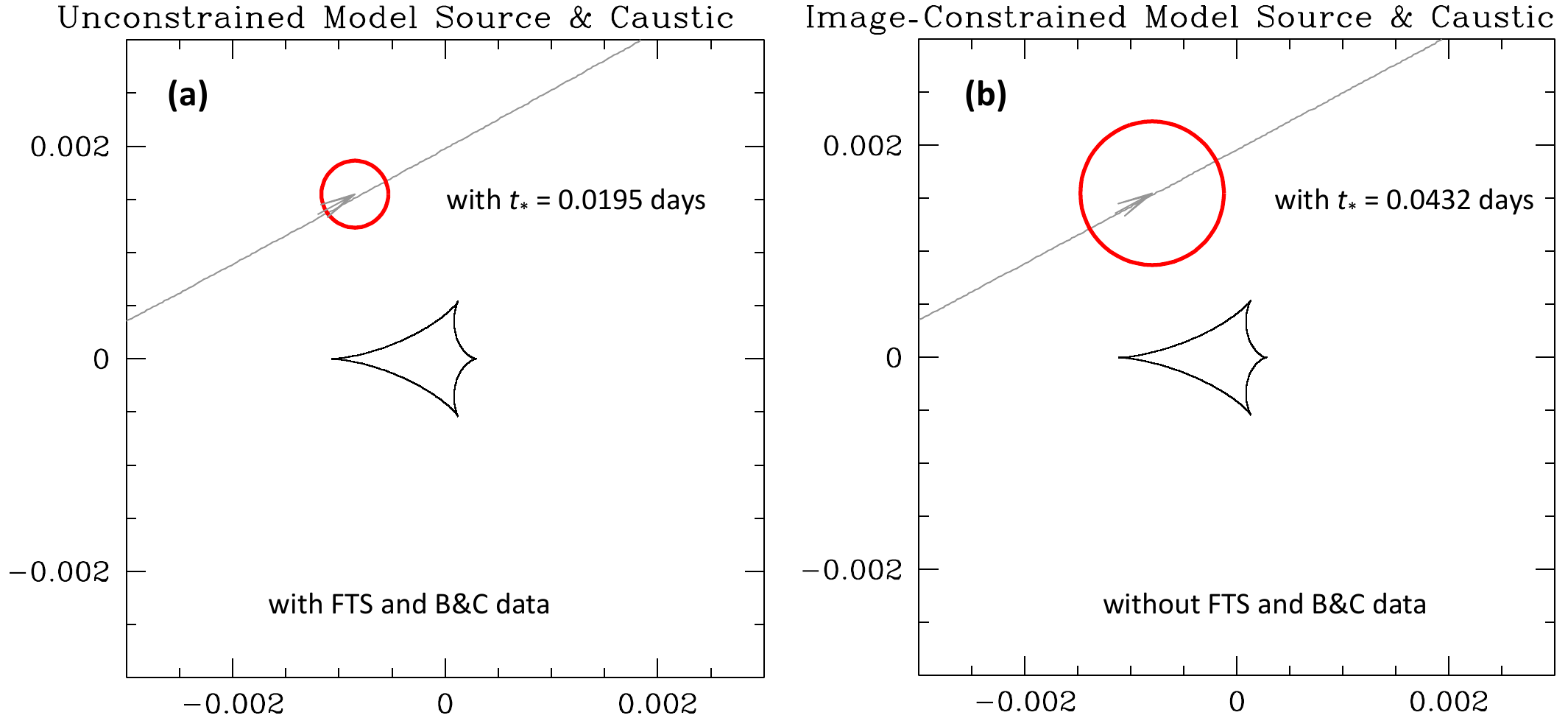}
\caption{Comparison of the source star trajectory, the central caustic, and the source star size (in red) for
the best fit unconstrained model, with (a) the best fit unconstrained model and (b) the best fit image-constrained
model with the FTS and B\&C data removed. The minimum separation of the source star limb from the central caustic
is 3.7 source radii the unconstrained model (a) and 1.0 source radii for the image-constrained model (b)
without the FTS and B\&C data.
}
\label{fig-caustic_comp}
\end{figure}

These constraints change the model results quite dramatically,, as Table~\ref{tab-fitcomp} indicates. The second column
of this table gives the fit results with no constraints, and the third column gives the with only the constraint on the 
$K$-band magnitude of the lens, $K_L = 17.786 \pm 0.185$, which yield essentially the same same conclusions
as the F15 paper. 
The fifth column gives the best fit model after removing two data sets that have apparent systematic errors
that we discuss in the next section.
However, the addition of the constraints on $\mubold_{\rm rel,H}$, and the lens and source
magnitudes increased the best fit source radius crossing time, $t_*$, by a factor of $\simgt 2$, over the best 
fit unconstrained model. This has a direct effect on the inferred angular Einstein radius, 
$\theta_{\rm E} = \theta_* t_{\rm E}/t_*$, which decreases by a factor of $\simgt 2$.

\begin{figure}
\epsscale{0.9}
\plotone{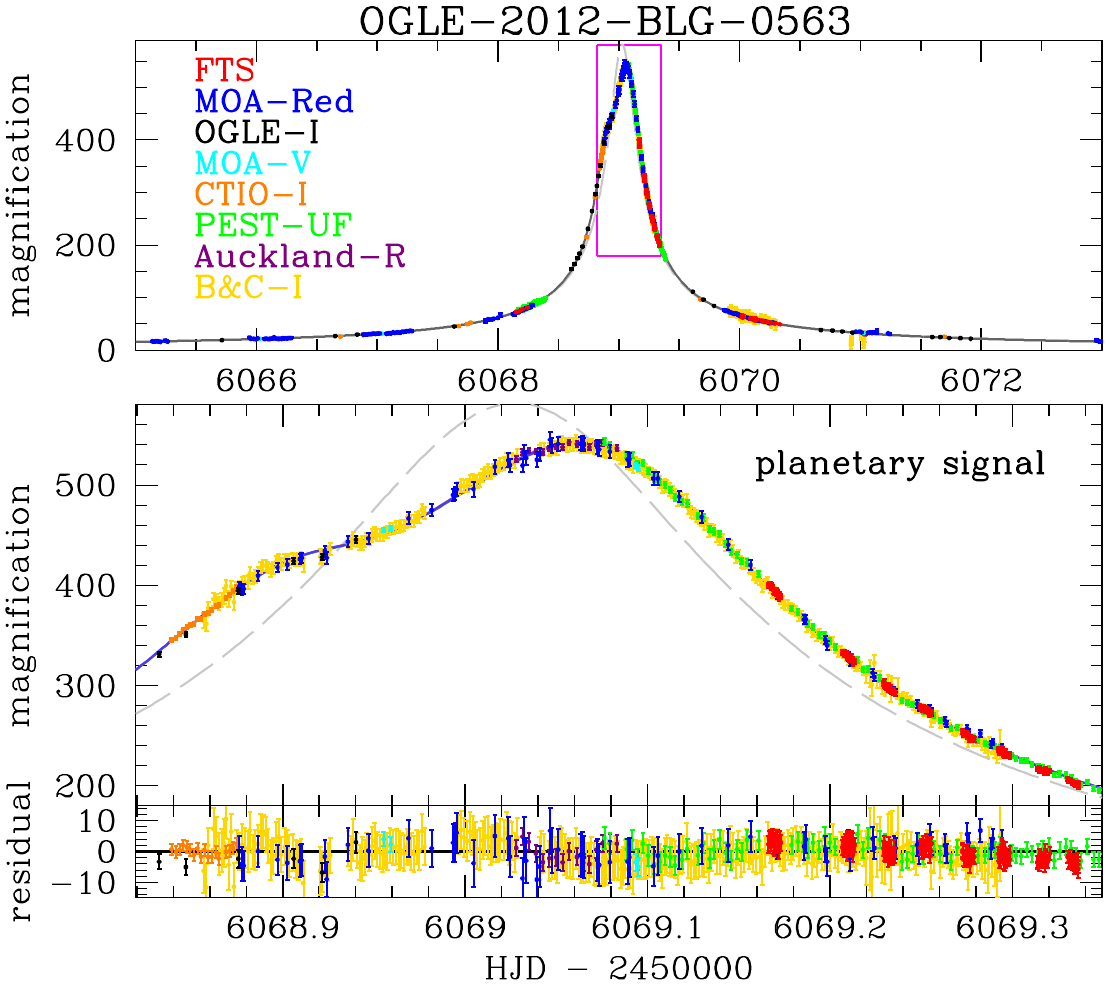}
\caption{The best fit light curve with the constraints from the high angular resolution follow-up data using the
full data set including the FTS and B\&C data, with the parameters given in the 4th column of Table~\ref{tab-fitcomp}.
}
\label{fig-lc_wFTS}
\end{figure}

\begin{figure}
\epsscale{0.98}
\plotone{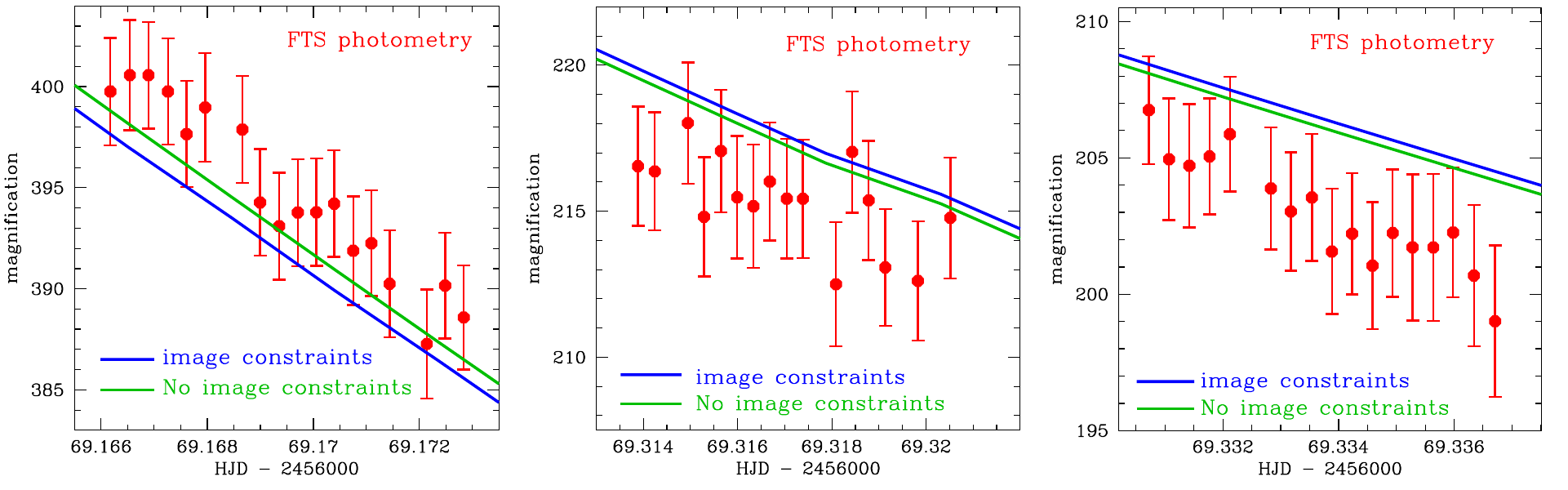}
\caption{A close-up comparison of the FTS data from the first and last two $\sim 10\,$minute observing intervals
on the night of the light curve peak. This data is compared to best fit models with and without the inclusion of constraints
from the high angular resolution images. The model light curves obtained with the constraints are shown in blue,
and the light curves without these constraints are shown in green.
The difference between the model curves is much smaller than the offset 
between the data and the models, that the FTS data is not a good fit to either model.
}
\label{fig-lc_error_FTS}
\end{figure}

\section{Systematic Errors in Light Curve Photometry}
\label{sec-syserror}

The existence of these systematic errors was foreshadowed by the inset plot in Figure 1 of F15 which
shows the central caustic, the source trajectory, and the source size for their best fit model. Figure~\ref{fig-caustic_comp}
shows a comparison of the central caustic and the source star trajectory and radius, for the best fit model with no
image constraints, including the FTS and B\&C data (a) and our final result with the high angular resolution image
constraints (b), which excludes the FTS and B\&C data. While the central caustic in panels (a) and (b) of 
Figure~\ref{fig-caustic_comp} look nearly identical, the source radius in panel (b), with the the constrained fit,
is twice the size of the source radius in panel (a). This is due to the larger source radius crossing time for the
image-constrained fit, as indicated in the second and fifth columns of Table~\ref{tab-fitcomp}. For the unconstrained fit
parameters illustrated in Figure~\ref{fig-caustic_comp}(a), the limb of the source star is always at least 3.7 source
star radii from the central central caustic. This is surprising because finite source effects typically are not detectable,
unless the limb of the source star passes within $\sim 1.0\,$source radii of the caustic, which is the minimum 
separation between the source star limb can central caustic for the image-constrained fit, illustrated in 
Figure~\ref{fig-caustic_comp}(b).

Table~\ref{tab-fitcomp} also lists the $\chi^2$ values for four of the data sets used in the light curve modeling
in order to show which data sets are driving the unconstrained fit towards very small source radius crossing times. 
For most microlensing events, it is the OGLE data that suffers the least degradation from systematic errors, due to
the excellent seeing at the OGLE observing site at the Las Campanas Observatory, and highly optimized
photometry code \citep{ogle4}, and we see that the $\chi^2$ of the OGLE-$I$ photometry is improved 
by the image constrained models. However, the FTS $\chi^2$ increases by $\Delta\chi^2 = 28.00$ and the
B\&C $\chi^2$ is increased by $\Delta\chi^2 = 13.86 $ when the image constraints are applied to the modeling.

In order to investigate this discrepancy between the high angular resolution imaging results and photometry from 
FTS and B\&C data, we take a closer look at the best image-constrained model, including all the data considered
in this paper, which is displayed in Figure~\ref{fig-lc_wFTS}. The FTS data is displayed in red, and it consists of
a series of continuous observations for periods of about 10 minutes at about a three hour interval. The residual
plot at the bottom of Figure~\ref{fig-lc_wFTS} shows that the FTS data gradually drifts from above the model
light curve to below the light curve over the time span of the FTS observations on the night of the light 
curve peak.

Figure~\ref{fig-lc_error_FTS} shows close-ups of the first and last two (which are fourth and fifth)
of these FTS observing periods, with the best fit model with and without the image constraints shown in 
blue and green, respectively. Note that neither model is a good fit to the FTS data, and the difference
between the blue and green curves is much smaller than the difference between the data and either curve.
Thus, it seems likely that the offset between the data and both models is due to a systematic error in the FTS 
data. As mentioned in section~\ref{sec-event} and illustrated in Figure~\ref{fig-OGLE_v_HST}, the 
photometry of this star is prone to systematic errors due to the proximity of a very 
bright star. We also note that it is more difficult to identify systematic errors in photometry from 
telescopes that are used to follow-up potential planetary microlensing events than it is for microlensing
survey telescopes, like the MOA and OGLE telescopes. This is because the survey telescopes include a large
number of observations in varying observing conditions without significant microlensing magnification. These can
be used to identify systematic errors and possibly remove them with detrending methods \citep{bennett12,bond17}.
In contrast, the data from follow-up telescopes often only include data with significant microlensing magnification,
and so it can be difficult to differentiate between systematic errors and higher order microlensing effects. Finally,
these follow-up telescopes often focus a much larger fraction of their observing time on candidate microlensing
events, so they may have many more observations than the survey telescopes. This can allow their systematic
errors to dominate some light curve model features over the survey telescopes (and OGLE, in particular), which
are generally less prone to systematic photometry errors.

\begin{deluxetable}{cccccc}
\tablecaption{Best Fit Model Parameters with $\mubold_{\rm rel,H}$ and Magnitude Constraints
                         \label{tab-Cmparams} }
\tablewidth{0pt}
\tablehead{
& \multicolumn{2}{c} {$u_0 < 0$} & \multicolumn{2}{c} {$u_0 > 0$} &  \\
\colhead{parameter}  & \colhead{$s<1$} & \colhead{$s> 1$} & \colhead{$s<1$} & \colhead{$s> 1$} &\colhead{MCMC averages}
}  
\startdata
$t_{\rm E}$ (days) & 63.768 & 63.818 & 63.773 & 64.166 & $63.8\pm 2.1$  \\   
$t_0$ (${\rm HJD}^\prime$) & 6069.0281 & 6068.8862 & 6069.0280 & 6068.8812 & $6068.96\pm  0.74$  \\
$u_0$ & -0.0017281 & -0.0005312 & 0.0017294 & 0.0004917 & $-0.00115\pm 0.00062$  \\
  & & &  \multicolumn{2}{c} {($u_0 > 0$)} & $0.00112\pm 0.00063$  \\
$s$ & 0.42452 & 2.36105 & 0.42416 & 2.38039 & $ 0.4230 \pm 0.0085$  \\
  & & &  \multicolumn{2}{c} {($s > 1$)} &  $2.376\pm 0.048$  \\
$\alpha$ (rad) & 0.49231 & 0.49249 & -0.49171 & -0.49185 & $ 0.4928\pm 0.0027$  \\
  & & &  \multicolumn{2}{c} {($u_0 > 0$)} &  $-0.4920\pm 0.0027$  \\
$q \times 10^{3}$ &1.3088 & 1.3070 & 1.3098 & 1.3273 & $1.331 \pm 0.085$  \\
$t_\ast$ (days) & 0.04319 & 0.04312 & 0.04305 & 0.04255 & $0.0431\pm 0.0019$ \\
$\pi_{\rm E,N}$ & 0.05821 & 0.05788 & 0.05814 & 0.05896 & $0.0583\pm 0.0060$ \\
$\pi_{\rm E,E}$ & 0.07955 & 0.08079 & 0.07971 & 0.08067 & $0.0801\pm 0.0046$\\
$D_{s}$ (kpc) & 8.4531 &8.2873 & 8.4462& 8.3923 & $8.48\pm 1.14$ \\
$\theta_{\rm E}$ (mas) & 0.6704  & 0.6698  & 0.6707 & 0.6752 & $0.645 \pm 0.17$ \\
fit $\chi^2$ & 1347.37 & 1347.97 & 1350.25 & 1350.63 &  \\
dof & $\sim 1460$ & $\sim 1460$ & $\sim 1460$ & $\sim 1460$ \\
\enddata \\
\vspace{0.15cm}
The $\pi_{\rm E}$ values are based on the inertial geocentric coordinate system that moves with \\
the Earth at $t_{\rm fix} = 6069\,{\rm HJD}^\prime$.
\end{deluxetable}

Also, the time period covered by the FTS data is also covered by a number of other data
sets including MOA and the Perth Exoplanet Survey Telescope (PEST), and these data sets do not favor the 
deviations from the light curve models seen by FTS. Therefore, we remove the FTS data from our
final light curve modeling. The B\&C data also had a $\chi^2$ increase when the image constraints are
added to the modeling, but the increase is smaller than the $\chi^2$ increase from the FTS data, while the
number of B\&C observations are larger. So, it is difficult to generate a figure similar to Figure~\ref{fig-lc_error_FTS} 
that offers a clear indication of systematic errors. However, B\&C  has also proved to be problematic for 
some previously analyzed events, so we drop this data as well. As columns 4 and 5 of Table~\ref{tab-fitcomp} 
show, the removal of the FTS and B\&C data has only a small effect on the best fit model parameters. We remove
it because it would be likely to induce unreasonably small error bars on the inferred parameters of the 
planetary microlens system.

\section{Lens Properties}
\label{sec-lens_prop}

Our final modeling of this event excludes the FTS and B\&C data, yielding the best fit model parameters
indicated in the fifth column of Table~\ref{tab-fitcomp}. We use the same image-constrained modeling 
method presented in \citet{bennett-moa379}. We used Gaussian constraints listed in Table~\ref{tab-mu-rel-mag}
for the weighted mean $\mu_{\rm rel,H,N}$ and $\mu_{\rm rel,H,E}$ values and the lens and source
magnitudes, except that there is no $K$-band light curve data to be constrained with the source $K$-band
measurement from {\sl Keck}. We also constrain the combined lens and source magnitudes
to be $I_{LS} = 19.186 \pm 0.080$ and $V_{LS} = 20.797 \pm 0.080$, as mentioned in Section~\ref{sec-HST}.
In addition to these nine constraints, we also constrain the
source distance, $D_S$ using a Galactic model prior from \citet{koshimoto_gal_mod}.

The model source magnitude values for these constraints come from our calibrations of the light curve 
photometry, but the lens magnitudes come from the empirical mass-luminosity relations presented in
\citet{bennett_moa291,bennett-ogle71}. In addition to these mass-luminosity relations, we must
also account for the dust extinction as a function of lens distance. At Galactic coordinates of
$l = 3.3120^\circ$ and $ b = -3.2518^\circ$, the extinction towards the lens system can be almost as
large as the extinction to the source star, unless the lens system is unusually close to the 
observer, as was the case for the F15 analysis.
We assume a dust scale height of $h_{\rm dust} = 0.10\pm 0.02\,$kpc \citep{drimmel}, so that the
extinction in the foreground of the lens is given by
\begin{equation}
A_{i,L} = {1-e^{-|D_L(\sin b)/h_{\rm dust}|}\over 1-e^{-|D_S (\sin b)/h_{\rm dust}|}} A_{i,S} \ ,
\label{eq-A_L}
\end{equation}
where the index $i$ refers to the passband: $I$, $V$, or $K$. 

\begin{deluxetable}{cccc}
\tablecaption{Measurement of Planetary System Parameters from the Lens Flux Constraints\label{tab-params}}
\tablewidth{0pt}
\tablehead{\colhead{parameter}&\colhead{units}&\colhead{values \& RMS}&\colhead{2-$\sigma$ range}}
\startdata
Angular Einstein Radius, $\theta_{\rm E}$&mas&$0.645\pm 0.017 $&0.611--0.681 \\
Geocentric lens-source relative proper motion, $\mu_{\rm rel, G}$&mas/yr&$3.694\pm  0.038$ & 3.619--3.770\\
Host star mass, $M_{\rm host}$&${\msun}$&$0.801\pm 0.033$ & 0.736--0.866\\
Planet mass, $m_{\rm pl}$&$M_{\rm Jup}$& $1.116\pm 0.087$ & 0.952--1.300\\
Host star - Planet 2D separation, $a_{\perp}$&AU&$4.77\pm 3.51$ & 1.24--9.90\\
Host star - Planet 2D sep. (close), $a_{\perp}$&AU&$1.50\pm 0.16$ & 1.20--1.81\\
Host star - Planet 2D sep. (wide), $a_{\perp}$&AU&$8.41\pm 0.87$ & 6.75--10.18\\
Host star - Planet 3D separation, $a_{3\rm d}$&AU & $3.76\pm 10.85$ & 1.33--28.06\\
Lens distance, $D_L$&kpc &$5.49\pm 0.56$& 4.45--6.64\\
Source distance, $D_S$ &kpc & $8.48\pm 1.14$& 6.41--10.88\\
\enddata
\end{deluxetable}

\begin{figure}
\epsscale{0.9}
\plotone{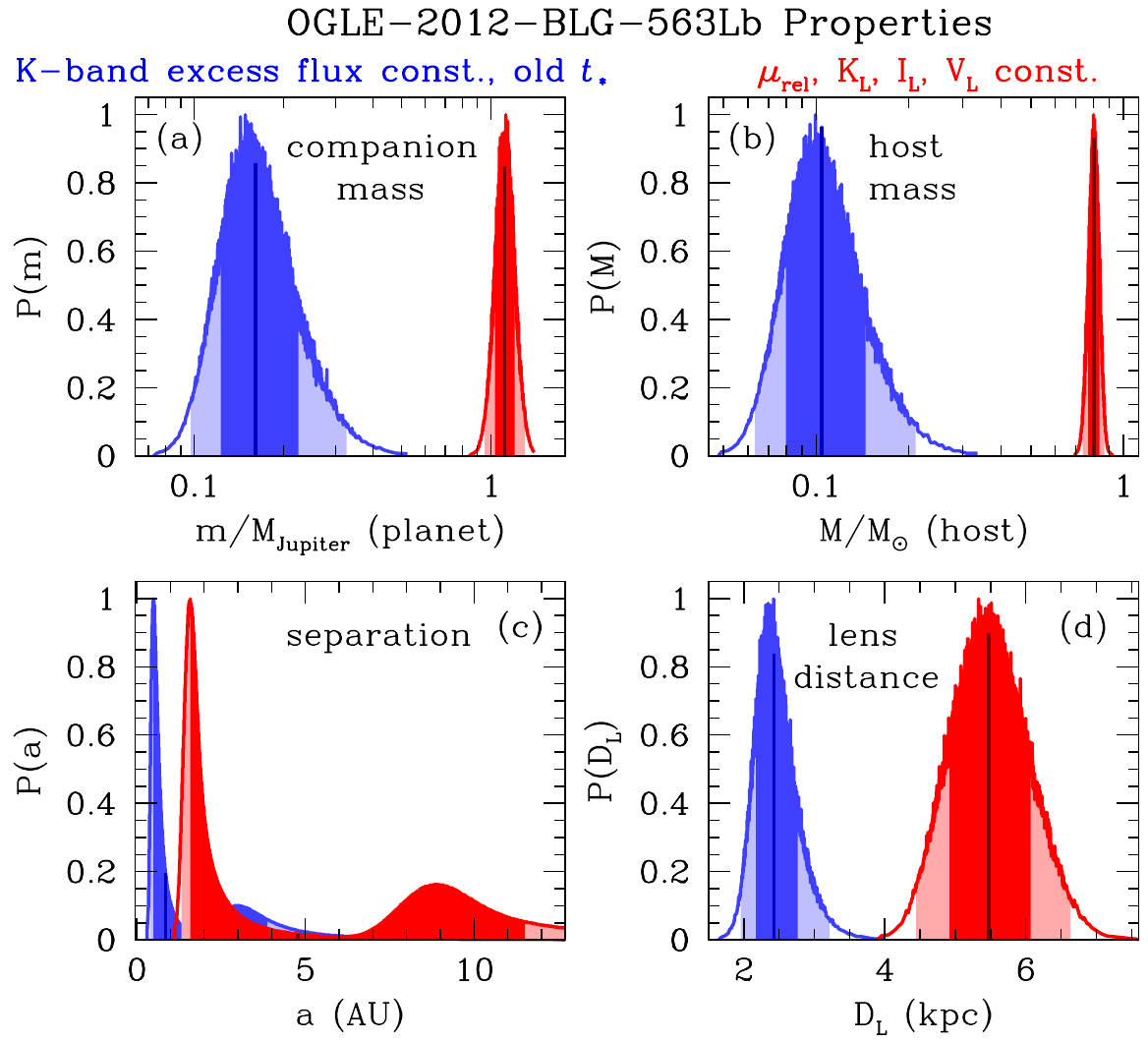}
\caption{Planetary system properties derived from out image-constrained modeling of
light curve data with constraints from high angular resolution follow-up constraints from {\sl Keck}
and {\sl Hubble} data in red, compared to the analysis of \citet{fukui15} in blue. The large discrepancy
is due to a systematic error in some of the light curve photometry that led to an underestimate of
the source radius crossing time, $t_*$, and therefore, a much larger angular Einstein radius, $\theta_{\rm E}$.
}
\label{fig-lens_prop}
\end{figure}

High magnification events like OGLE-2012-BLG-0563
are subject to the close-wide degeneracy \citep{dominik99} due to the fact the that the model parameter
transformation $s \rightarrow 1/s$ leaves the characteristics of the central caustic almost unchanged. A more
general version of this degeneracy has been described by \citet{zhang22}. Galactic bulge microlensing events 
observed toward the Galactic bulge with microlensing parallax signals are also subject to the ecliptic degeneracy
\citep{poindexter05}, which is exact for events in the ecliptic plane. This degeneracy involves replacing
a binary lens system with its mirror image, and it is the orbital motion of the Earth, which is detected via the 
microlensing parallax effect, that breaks the mirror symmetry. This ecliptic degeneracy results in a sign
change for the for the $u_0$ and $\alpha$ parameters. These two degeneracies lead to the four degenerate 
models. The parameters for the best fit models in each of these degenerate categories are
given in Table~\ref{tab-Cmparams}. This table also shows the results of our Markov Chain Monte Carlo
calculations over all the degenerate model. The parameters are quite similar for all four degenerate, except
for the parameters directly affected by the degeneracies: $u_0$, $\alpha$, and $s$. The parameters 
$u_0$ and $\alpha$ are not generally considered to be physically interesting because these just depend 
on the detailed alignment between the lens and source and the orientation of the lens system (which is 
thought to be random). For events with $s\approx 1$, the $s \leftrightarrow 1/s$ degeneracy has little effect
on the inferred physical parameters \citep[e.g.][]{batista15,bennett16,bennett-moa379,aparna18}, because the 
uncertainty due to the unobserved line-of-sight separation is larger than the difference between the close
and wide solutions. However, for OGLE-2012-BLG-0563, the 
close and wide solutions have projected separations that differ by a factor of $5.62\pm 0.11$, so the 
orbital separations predicted by the close and wide solutions have very little overlap, as shown by 
the red curve in Figure~\ref{fig-lens_prop}(c) and the host star and planet 2D separation values listed in
Table~\ref{tab-params}. This indicates that the host star is a K dwarf with a mass of 
$M_{\rm host} = 0.801 \pm 0.033\msun$, at a distance of $D_L = 5.49\pm 0.56\,$kpc, orbited by a 
planet of mass $m_{\rm pl} = 1.116\pm 0.087 M_{\rm Jup}$. The separation of the planet from its host star
is less certain, due to the close-wide light curve model degeneracy. The 2-dimensional planet-host star
separation is $a_\perp = 1.50\pm 0.16$ for the close solution and $a_\perp = 8.41\pm 0.87$ for the 
wide solution, with a 2-$\sigma$ range of 1.24--9.90\,AU, when both solutions are considered. If we assume 
a random orientation of the planetary lens system, we find a 2-$\sigma$ range of 1.33--28.06\,AU,
for the 3-dimensional separation.

Figure~\ref{fig-lens_prop} also compares the planetary system properties found by our analysis to the 
results of a similar analysis using only the constraints considered in the F15 paper.  However,
this comparison analysis using the full data set, including the FTS and B\&C data with only the
$K_L$ magnitude constraint yields a slightly different result that the one reported by F15.
Figure 8 of F15 indicates a double-peaked host mass and distance distribution with a sharp peak
at $M_{\rm host} \approx 0.13 \msun$, $D_L \approx 0.5\,$kpc and a broader peak at 
$M_{\rm host} \approx 0.43 \msun$, $D_L \approx 1.5\,$kpc. Part of the reason for this difference
is illustrated by Figure 7 of F15, which shows that the mass-distance relations from the
angular Einstein radius, $\theta_{\rm E}$ and the lens $K$ band magnitude, $K_L$, are nearly parallel 
and have a significant overlap for $0.4\,{\rm kpc} < D_L < 2\,$kpc. The F15 analysis also assumes
that the interstellar extinction is uniformly distributed along the line-of-sight to the source, which ignores
measures that show the dust scale height in the Galactic disk is much lower than the scale height
of the stars \citep{drimmel} as represented by equation~\ref{eq-A_L}. Thus, F15 underestimate 
the extinction toward possible host stars at smaller distances (\ie\ $D_L \simlt 3\,$kpc). A more realistic
extinction estimate would increase the $\theta_{\rm E}$ mass-distance relation in Figure 7 of F15.
Other differences between our analysis and that of  F15 are that we have used re-reduced 
photometry using an improved method \citep{bond17} and we have used an empirical mass-luminosity 
relation.

\begin{figure}
\epsscale{0.8}
\plotone{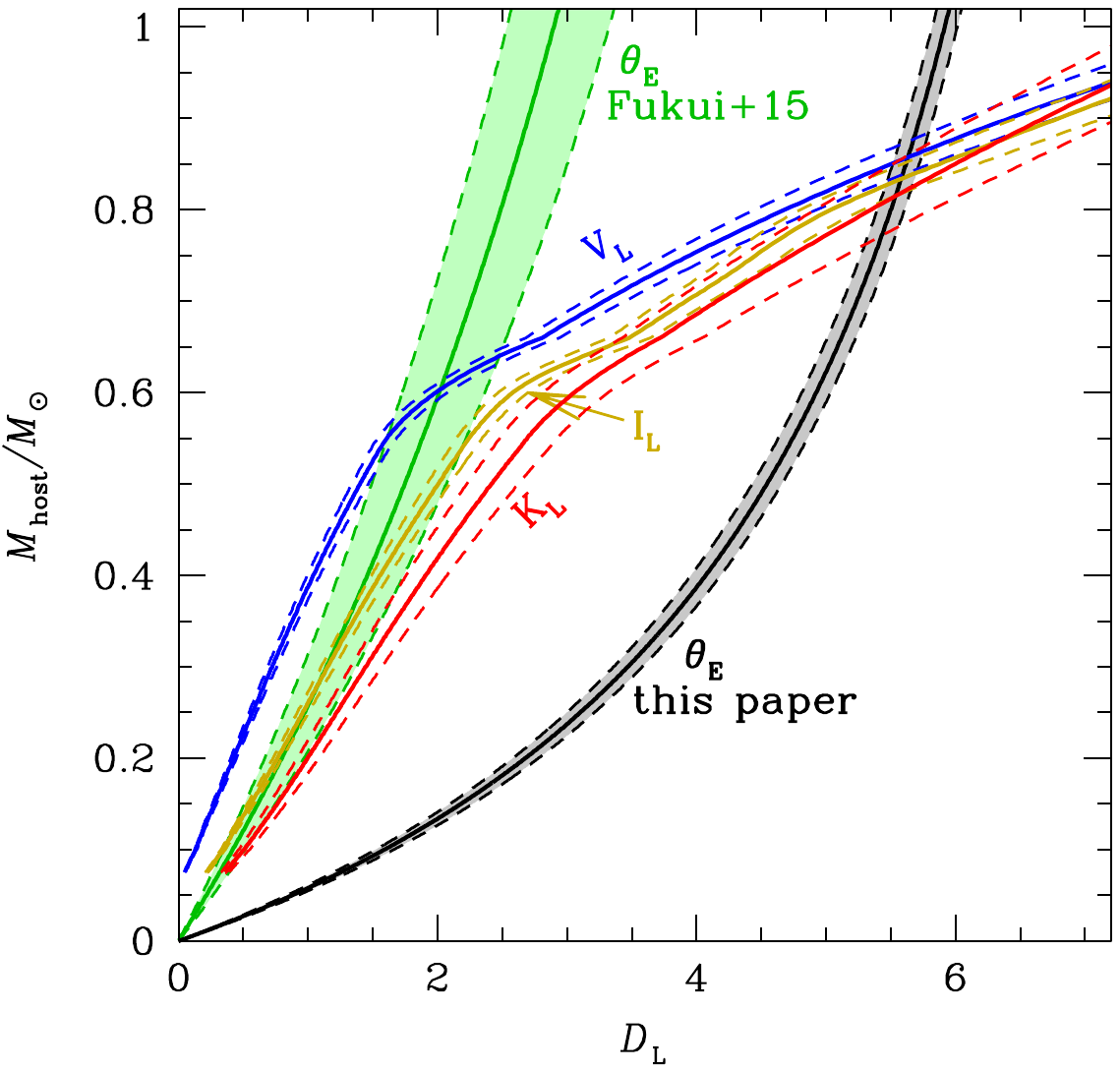}
\caption{Mass-distance relations from {\sl Hubble} and {\sl Keck} host star brightness
measurements and angular Einstein radius, $\theta_{\rm E}$, determinations from \citet{fukui15} (in green)
and this paper (in black and grey). The lens star brightness mass-distance constraints in the $V$, $I$, and $K$ bands
are presented in blue, gold, and red, respectively, and the dashed lines indicate 1$\sigma$ error bars.
Our smaller $\theta_{\rm E} = 0.645\pm 0.017\,$mas value intersects all three host magnitude 
mass-distance constraints at $D_L \approx 5.5\,$kpc, implying a $M_{\rm host} \approx 0.8\msun$, but
the \citet{fukui15} value, $\theta_{\rm E} = 1.36\pm 0.14\,$mas, intersects the lens star brightness mass-distance constraints
at different masses and distances for each passband, implying a very nearby, low-mass host star for 
the $K$ band constraint, which is similar to the \citet{fukui15} value.
}
\label{fig-mass_dist}
\end{figure}

It may seem odd that a reduction in the inferred angular Einstein radius would lead to a significant increase
in the implied mass of the lens system, because the $\theta_{\rm E}$ mass-distance relation,
\begin{equation}
M_L = {c^2\over 4G} \theta_{\rm E}^2 {D_S D_L\over D_S - D_L}  \ ,
\label{eq-m_thetaE}
\end{equation}
implies that the $M_L$ increases quadratically with $\theta_{\rm E}$. Of course, the answer is that the 
lens mass depends on both its distance, $D_L$, and  $\theta_{\rm E}$. Figure~\ref{fig-mass_dist} 
displays a comparison of the effects of the $\theta_{\rm E}$ determinations by F15 and
our analysis with the lens magnitude constraints in the $V$, $I$, and $K$ passbands. The $K$ band
lens magnitude constraint only intersects with the F15 $\theta_{\rm E}$ curve at very small distances
and very low host masses, but the $I$ and and especially the $V$ band curves intersect with the 
F15 $\theta_{\rm E}$ curve at larger distances and masses. This is an indication of the systematic
photometry error that led to the overestimate of the F15 $\theta_{\rm E}$ value. Note that 
Figure~\ref{fig-mass_dist}  is a simplified presentation of the results. It includes the uncertainties
on the measured host star magnitudes, and the average error bars for the $\theta_{\rm E}$, but the
results presented in Figure~\ref{fig-lens_prop} and Table~\ref{tab-params} also include the effects of
correlations with the other event model parameters from the MCMC analysis.

\section{Conclusions}
\label{sec-conclude}

Our analysis of the 2018  {\sl Hubble} images for planetary microlensing event OGLE-2012-BLG-0563 finds
a substantially larger mass and distance for the planetary system with host star and planet masses of 
$M_{\rm host} = 0.801\pm 0.033\msun$ and $M_{\rm planet} = 1.116 \pm 0.087 M_{\rm Jupiter}$, respectively.
They are located at a distance of towards the Galactic bulge, with a projected star-planet separation of
$1.50\pm 0.16\,$AU or $8.41\pm 0.87\,$AU, depending on whether the degenerate close or wide model is
correct. Our light curve analysis used the image-constrained modeling version \citep{bennett-moa379}
of the \texttt{eesunhong} \citep{bennett96,bennett-himag} light curve modeling code. This modeling revealed
a discrepancy between the source radius crossing time, $t_*$, favored by some of the light curve data from 
microlensing follow-up observations and the measured lens-source relative proper motion, $\mubold_{\rm rel,H}$,
measured independently by {\sl Hubble} images in the $V$ and $I$ bands, as well as {\sl Keck} adaptive 
optics images in the $K$ band \citep{aparna24}. This suspect ground-based photometry from the
Faulkes South Telescope (FTS) to a lesser extent the MJUO B\&C Telescope was investigated
and found to be due to a minor $\chi^2$ improvement, that only slightly reduced a discrepancy from
every model light curve, including the light curve with the small $t_*$ favored by this data. Also, the $\chi^2$ values
for the data sets that are historically the most reliable, such as OGLE, were improved when the high angular
resolution imaging constraints were added to the modeling.

The analysis of this event also indicates the importance of redundant methods for determining that mass and
distances for planetary systems found by microlensing. In this case, have mass-distance relations from observations
in three passbands, redundant determinations of $\theta_{\rm E}$ from light curve measurements of the source
radius crossing time, $t_*$, and measurements of the lens-source relative proper motion, $\mubold_{\rm rel,H}$,
from the high angular resolution follow-up observations. These $\theta_{\rm E}$ values also provide a mass-distance
relation, as indicated in Figure~\ref{fig-mass_dist} and equation~\ref{eq-m_thetaE}. Also, the image-constrained modeling
enforces consistency between the implied lens mass from its magnitudes, the $\mubold_{\rm rel,H}$ value, and the
microlensing parallax light curve parameters, $\piEbold$. These redundant constraints on the physical 
parameters of the planetary microlensing system allowed the systematic errors in the light curve photometry to be quickly
identified and corrected.

We anticipate that this image-constrained modeling method will be very useful for analyzing the microlens planetary
systems discovered by the exoplanet microlensing survey of the {\sl Nancy Grace Roman Space Telescope} 
\citep{bennett_MPF,bennett18_wfirst,WFIRST_AFTA,penny19}. This survey, known as the Roman Galactic Exoplanet
Survey (RGES), will use the same methods to determine the masses of and distance to planetary microlensing systems
as we have done with high angular resolution follow-up data from {\sl Keck} and {\sl Hubble}, except that the 
high angular resolution RGES images will be employed to determine the image constraints. RGES photometry
will generally not be affected by the same types of systematic errors that we have discovered in the ground-based
photometry used for the \citet{suzuki16} statistical sample of planetary microlensing events. However, the {\sl Roman
Space Telescope} employs a new generation of infrared detectors. While these detectors have improved sensitivity
over previous generations of infrared detector \citep{mosby20,teledyne_H4RG10}, their improved sensitivity is likely
to reveal subtle systematic errors that were not apparent in data from previous generations of detectors. We expect
that the image-constrained light curve modeling method will enable any such systematic errors to be
efficiently identified.

\acknowledgments 
This paper is based in part on observations made with the NASA/ESA {\sl Hubble Space Telescope}, 
which is operated by the Association of Universities for Research in Astronomy, Inc., under NASA contract NAS 5-26555. 
These observations are associated with program GO-15455.
The {\sl Keck} Telescope observations and analysis were supported by a NASA {\sl Keck} PI Data Award, administered by the 
NASA Exoplanet Science Institute. Data presented herein were obtained at the W. M. {\sl Keck} Observatory from telescope 
time allocated to the National Aeronautics and Space Administration through the agency's scientific partnership 
with the California Institute of Technology and the University of California. The Observatory was made possible by 
the generous financial support of the W. M. {\sl Keck} Foundation. 
DPB, AB, NK, SKT, and AV were also supported by 
NASA through grants 80NSSC20K0886, 80GSFC21M0002 and 80NSSC24M0022. 
Some of this research has made use of the NASA Exoplanet Archive, which is operated by the California Institute of 
Technology, under contract with the National Aeronautics and Space Administration under the Exoplanet Exploration Program.
This work was supported by the University of Tasmania through the UTASFoundation and the endowed Warren Chair 
in Astronomy and the ANR COLD-WORLDS (ANR-18-CE31-0002).
This work was also supported by JSPS Core-to-Core Program JPJSCCA20210003. 

\end{document}